# On the Profitability of Optimal Mean Reversion Trading Strategies

Peng Huang

IEOR Department, Columbia University, New York, NY 10027

peng.huang@columbia.edu

Tianxiang Wang

IEOR Department, Columbia University, New York, NY 10027

tw2517@columbia.edu

This Version: January 2016

**WORKING PAPER**

# On the Profitability of Optimal Mean Reversion Trading Strategies

**ABSTRACT**

We study the profitability of optimal mean reversion trading strategies in the US equity market. Different from regular pair trading practice, we apply maximum likelihood method to construct the optimal static pairs trading portfolio that best fits the Ornstein-Uhlenbeck process, and rigorously estimate the parameters. Therefore, we ensure that our portfolios match the mean-reverting process before trading. We then generate contrarian trading signals using the model parameters. We also optimize the thresholds and the length of in-sample period by multiple tests. In nine good pair examples, we can see that our pairs exhibit high Sharpe ratio (above 1.9) over the in-sample period and out-of-sample period. In particular, Crown Castle International Corp. (CCI) and HCP, Inc. (HCP) achieve a Sharpe ratio of 2.326 during in-sample period and a Sharpe ratio of 2.425 in out-of-sample test. Crown Castle International Corp. (CCI) and Realty Income Corporation (O) achieve a Sharpe ratio of 2.405 and 2.903 respectively during in-sample period and out-of-sample period.

**Keywords**: maximum likelihood estimation, Ornstein-Uhlenbeck process, mean-reversion trading

**JEL Classification**: C41, G11, G12

# 1 Introduction

Mean reversion trading strategies are widely used in industry. However, not all strategies ensure that the portfolio value matches mean-reverting process before enacting the strategies. For example, Avellaneda and Lee(2009) model residuals of stock return as Ornstein-Uhlenbeck (OU) process and estimate the parameters before proving that the residuals exhibit the mean-reverting process. Inspired by Leung and Li (2015) and by Leung and Li (2016), we use maximum likelihood method to estimate the parameters and construct the optimal static pairs trading portfolio. For any pair, we construct the portfolio that best fits the OU process.

We pair a wide range of stocks to find the pairs that exhibit the OU process. We then trade pair portfolios according to our trading rules. Similar to other mean reversion trading strategies, we sell when the portfolio value is abnormally high and buy when the value is extremely low. However, we want to optimize the entry/exit trading signal. Therefore, we test a series of thresholds. We derive the Sharpe ratio, annualized return, maximum drawdown, trading frequency, trade range, return per trade for different thresholds. We select the best threshold for our out-of-sample test. Besides, considering the different industry cycle and economy cycle, we aim to find the proper in-sample period for each pair. Therefore, we test 4 in-sample periods but make the out-of-sample period constant. At last, we find 9 pairs which exhibit high return over the in-sample period and out-of-sample period. All pairs achieve above 1.9 Sharpe ratio in in-sample and out-of-sample test. Besides, we find that when the gap between entry signal and close signal is narrow, the trading frequency will be high. When the exit signal is near zero, the time we hold the portfolio is long. These two results are easy to understand, because assuming the gap is narrow, trading signal will achieve the thresholds more frequently. And when the close signal is near zero, we are less likely to close the trade. In the paper, we summarize the statistics results for nine pairs and display the detailed results for two pairs.

The rest of the paper is structured as follows. We introduce our methodology in Section 2 followed by a discussion on our pair examples in Section 3.

# 2 Methodology

Our implementation of pair trading has two stages, i.e. in-sample period and out-of-sample period. First, we form pairs using Maximum likelihood Estimation over the in-sample period. We then test our trading strategies under different thresholds and find the best entry/exit signal. Second, in our out-of-sample test, we backtest our pairs from 2014.12.23 to 2015.11.10(200 days). Noteworthy, we try in-sample period of different length (880 days, or 628 days, or 376 days, or 124 days) to find the most predictive one for out-of-sample test.

## 2.1 Pair Formation

We select five sectors from 156 sub industries listed in Global Industry Classification Standard (GICS), i.e. Banks, Internet Software & Services, Diversified Financial Services, REITs, and Health Care Distribution & Services. Then we pair any two stocks from the same sector. Noteworthy, all our stocks are in the S&P 500.

Our portfolio is to long $\alpha$ shares of a risky asset $S^{(1)}$ and short $\beta$ shares of another risky asset $S^{(2)}$:

$$\text{Portfolio: } x_t^{\alpha,\beta} = \alpha S_t^{(1)} - \beta S_t^{(2)}$$

To find the best (α, β) strategy, we referred to the methods provided by Leung and Li (2015). We used the maximum likelihood estimation (MLE) to fit the observed portfolio values to an OU process and determine the model parameters. Then we obtain the $(α^*, β^*)$ with highest average log-likelihood.

In detail, assuming we invest A dollar(s) in asset $S^{(1)}$, so $α = A/S_0^{(1)}$. Similarly, we short $β = B/S_0^{(2)}$ shares in $S^{(2)}$, in which B/A=0.001,0.002,…,1. Without loss of generality, we set A=1. Therefore, for each pair, we can keep α constant while B vary from 0,001 to 1. We apply maximum likelihood estimation (MLE) to fit the observed portfolio values $x_t^{α,β}$ to OU process and determine the model parameters. We then determine the B which shows a highest log-likelihood.

## 2.2 Signal Generation

Similar to the Avellaneda and Lee(2009), we trade only when we think that we detect an anomalous excursion. We consider an estimation window of 60 business days. Every trading day, we use data during estimation window to get the parameters of the OU process.

OU process:
$$dx_t = μ(θ − x_t)d_t + σdB_t$$

The equilibrium variance is
$$σ_{eq,i} = σ_i/\sqrt{2μ}$$

Accordingly, we define the dimensionless variable(s-score) for pair i:
$$s_i = \frac{x_{i,t} − θ_i}{σ_{eq,i}}$$

After we use the maximum likelihood method, we can derive the estimations of all the model parameters. Thus, we can calculate the trading signal $s_i$ every trading day.

Our mean-reversion trading rule is

$$\text{Buy to open if } s_i < −S_o$$
$$\text{Sell to open if } s_i > S_o$$
$$\text{Close short position if } s_i < S_c$$
$$\text{Close long position if } s_i > −S_c$$

We test 20 $S_o$ and 20 $S_c$. In total, we have 400 pairs of thresholds.

$$S_o = 1, 1.05, 1.1, …, 2$$
$$S_c = 0, 0.05, 0.1 …, 1.$$

## 2.3 Index Computation

For any pair, we calculate five indexes over the in-sample period and out-of-sample period, i.e. the annualized return, Sharpe ratio, the average time each trade last, the annualized trade frequency, and the return per trade.

We calculated the daily return $ret_{t+1}$ as:
$$ret_{t+1} = \left(\frac{x_{t+1}−x_t}{Cost_t}\right) * Position_t,$$

where $x_{t+1}$ is the portfolio value at day t + 1. $Cost_t = \alpha S_t^{(1)} + \beta S_t^{(2)}$, referring to the money we invest to long and short the stocks at day t. If $Position_t$ is -1, it refers that we short the portfolio at day t; if $Position_t$ is 1, it refers that we long the portfolio at day t; if $Position_t$ is 0, it refers that we close trade at day t.

Annualized return over the sample period is defined as:

$$\text{Annualized\_Return} = \left(\prod_{t=1}^{k}(1+\text{ret}_t)\right)^{\frac{252}{k}},$$

where k is the length of our sample. For example, assuming our in-sample period lasts 124 days, k equals to 124.

Sharpe ratio is defined as:

$$\text{Sharpe\_Ratio} = \frac{\text{Annualized\_Return}}{\text{Std(ret)}*\sqrt{252}},$$

where Std(ret) is the standard deviation of daily return.

To know how long we hold the portfolio, we define the variable TRange. For any complete trade in our sample period, we calculate the time between entering and exiting.

$$\text{TRange} = \frac{\sum_{i=1}^{i=n}(\text{exit\_time}_i - \text{entry\_time}_i)}{n},$$

In the above equation, $\text{exit\_time}_i$ is the time we close trade i, $\text{entry\_time}_i$ is the time we enter into trade i. n is the number of complete trade we enter in our test period.

Assuming that there are 252 trading days every year, we define the annualized Trade Frequency as:

$$\text{TFreq} = \frac{n*252}{k},$$

Next, we derive the average annualized return of per trade as:

$$\text{RetPerT} = \frac{\text{anualized\_return}}{\text{TFreq}}.$$

# 3 Pair Example
## 3.1 Summary statistics for nine good pairs

In the Table 1, we summarize the Sharpe ratio, annualized return, maximum drawdown, trading frequency, trade range, return per trade for the nine good pairs. We also show the parameters from OU process.

From this table, we can see all our pairs achieve above 1.9 Sharpe ratio in in-sample and out-of-sample tests.

In the Section3.2, we will show the results of two good pairs in detail.

| Sector | GICS Sector | Financials | Consumer Discretionary | Financials | Financials | Financials | Financials | Info Tech | Financials | Gold |
|---|---|---|---|---|---|---|---|---|---|---|
| | GICS Sub Industry | REITs | Hotels, Resorts & Cruise Lines | Banks | REITs | REITs | REITs | Internet Software & Services | REITs | Gold |
| Portfolio | Long(/$) | 1 CCI | 1 PCL | 1 PNC | 1 EQR | 1 CCI | 1 CCI | 1 EBAY | 1 CCI | 1 GLD |
| | Short(/$) | 0.173 HCP | 0.347 GGP | 0.793 WFC | 0.513 O | 0.336 SPG | 0.273 HCN | 0.381 INTU | 0.367 O | 0.515 GDX |
| Parameters | $\hat{\theta}$ | 0.9319 | 0.5213 | 0.2455 | 1.2082 | 0.6296 | 0.8164 | 0.5780 | 0.7204 | 0.6211 |
| | $\hat{\mu}$ | 9.9715 | 2.3036 | 15.7538 | 0.3042 | 14.5032 | 11.0183 | 9.2776 | 10.7316 | 6.5194 |
| | $\hat{\sigma}$ | 0.1969 | 0.1267 | 0.1466 | 0.1482 | 0.1753 | 0.2025 | 0.2093 | 0.1960 | 0.1278 |
| | $\hat{l}$ | 2.9904 | 3.4160 | 3.2970 | 3.2553 | 3.1158 | 2.9643 | 2.9278 | 2.9963 | 3.4158 |
| In sample | Length(days) | 280 | 220 | 440 | 260 | 200 | 380 | 220 | 380 | 840 |
| | SR | 2.326 | 2.353 | 2.265 | 2.796 | 2.231 | 2.246 | 2.276 | 2.405 | 1.902 |
| | AR | 0.236 | 0.075 | 0.097 | 0.07 | 0.195 | 0.232 | 0.089 | 0.217 | 0.076 |
| | DD | -0.06 | -0.012 | -0.023 | -0.012 | -0.039 | -0.054 | -0.022 | -0.06 | -0.02 |
| | TF | 7.20 | 8.02 | 21.19 | 4.85 | 12.60 | 9.95 | 6.88 | 8.62 | 4.50 |
| | TR | 15 | 8 | 4 | 5 | 9 | 11 | 5 | 14 | 15 |
| | RPT | 0.034 | 0.009 | 0.005 | 0.017 | 0.016 | 0.026 | 0.015 | 0.016 | 0.019 |
| Out of Sample | Length(days) | 200 | 200 | 200 | 200 | 200 | 200 | 200 | 200 | 200 |
| | SR | 2.425 | 2.292 | 2.7 | 2.679 | 2.073 | 2.495 | 1.933 | 2.903 | 2.298 |
| | AR | 0.215 | 0.127 | 0.067 | 0.121 | 0.141 | 0.185 | 0.129 | 0.243 | 0.076 |
| | DD | -0.033 | -0.022 | -0.009 | -0.013 | -0.022 | -0.032 | -0.053 | -0.033 | -0.011 |
| | TF | 10.08 | 11.34 | 16.38 | 10.08 | 17.64 | 16.38 | 10.08 | 15.12 | 3.78 |
| | TR | 13 | 7 | 3 | 5 | 4 | 5 | 5 | 9 | 6 |
| | RPT | 0.021 | 0.011 | 0.004 | 0.012 | 0.008 | 0.011 | 0.013 | 0.016 | 0.02 |

SR refers to the Sharpe ratio; AR refers to annualized return; DD refers to maximum drawdown; TF refers to annualized trading frequency; TR refers to average trading range; RPT refers to return per trade.

Table 1

## 3.1 CCI vs HCP

We construct a portfolio by holding $1 Crown Castle International Corp. (CCI ) and shorting $0.173 stock HCP, Inc. (HCP). The pair of stocks are selected from REITs sector. We simulate the OU process using estimated parameters. We show the parameters in Table 2. As we can see, the simulated process's parameters is near to the empirical one.

From Figure 1, we can see that this portfolio shows a high level of mean reversion. In Figure 2, we show the cumulative returns over the in-sample period under different thresholds.

From Figure 3 to Figure 7, we show the heat map of Sharpe ratio, annualized return, annualized trading frequency, annualized trading range, return per trade, under different thresholds.

We select thresholds under which we achieve highest Sharpe ratio during in-sample period. Figure 8 and Figure 9 are results under the best thresholds ($S_o = 1.5, S_c = 0$).

Figure 8 records the threshold, portfolio value, compound return, s-score, daily return over the in-sample period. Figure 9 is similar to Figure 8, except that the Figure 9 records the data over the out-of-sample period.

As we mentioned in Section 2, we try four different in-sample periods, i.e. 880 days, or 628 days, or 376 days, or 124 days. However, we don't show all the results under different in-sample period. To be concise, we only show the results during the best in-sample period. We select the sample period which is most predictive for out-of-sample test. As for the trading range and trading frequency pattern, from Figure 5 and Figure 6, we found that when the entry signal is close to exit signal, the trading frequency will be high. When exit signal is near zero, the trading range will last long.

|  | Price | $\hat{\theta}$ | $\hat{\mu}$ | $\hat{\sigma}$ | $\hat{l}$ |
|---|---|---|---|---|---|
| CCI-HCP | empirical | 0.9319 | 9.9715 | 0.1969 | 2.9904 |
|  | simulated | 0.9396 | 9.5228 | 0.1979 | 2.9847 |

Table 2

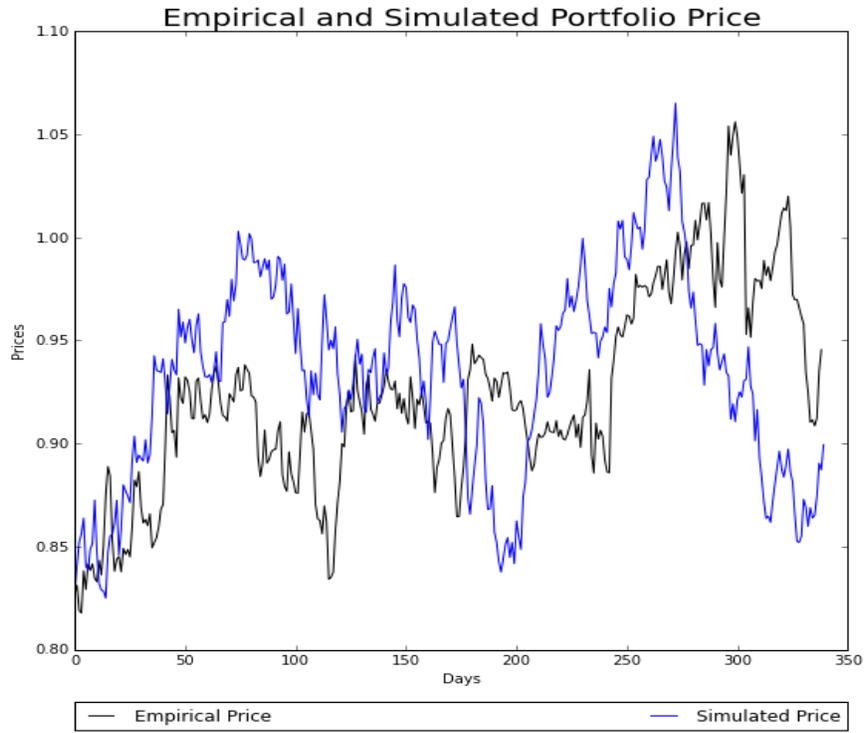

Figure 1

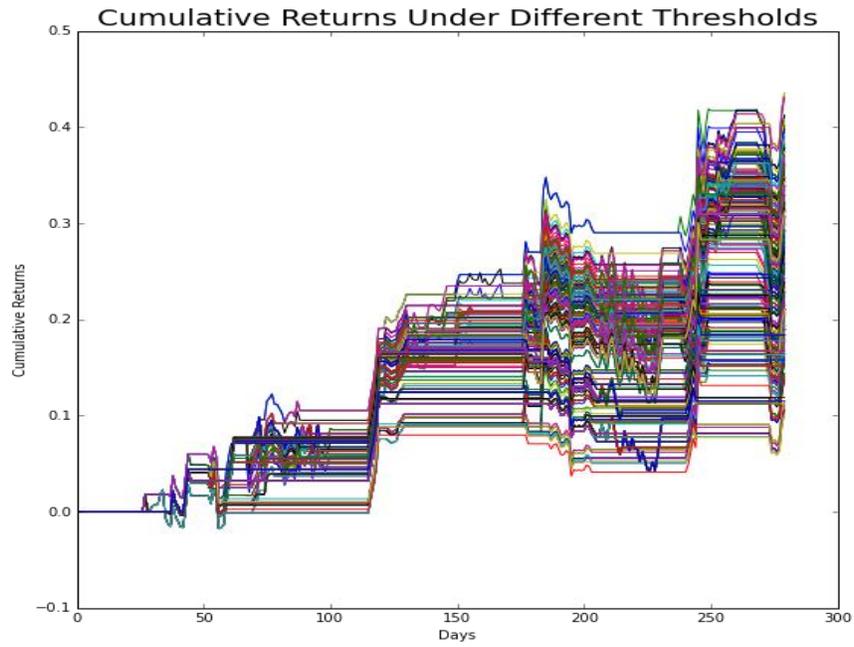

Figure 2

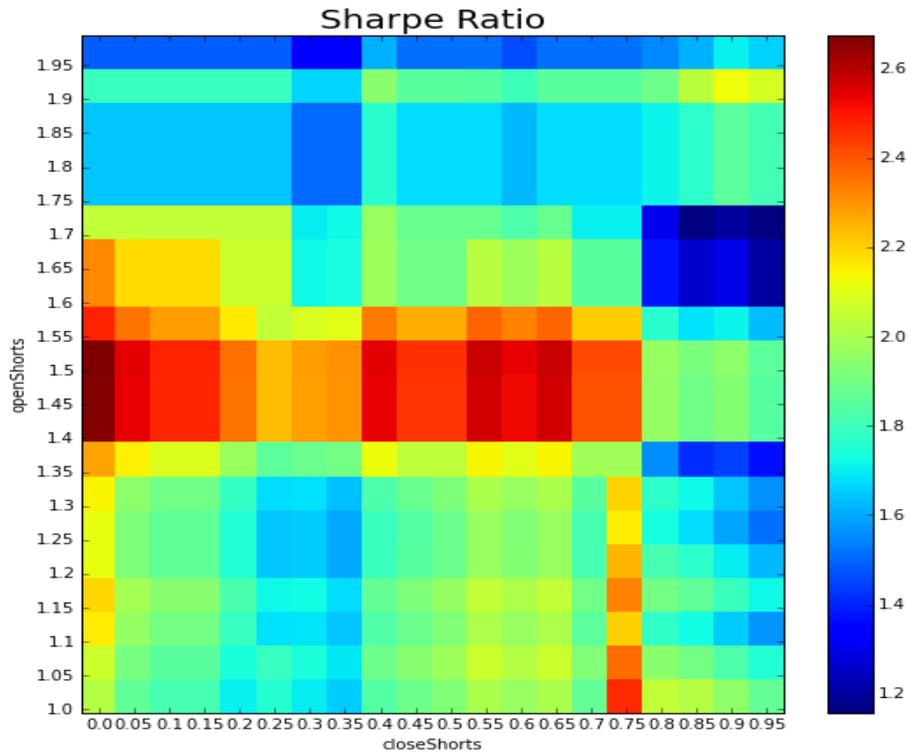

Figure 3

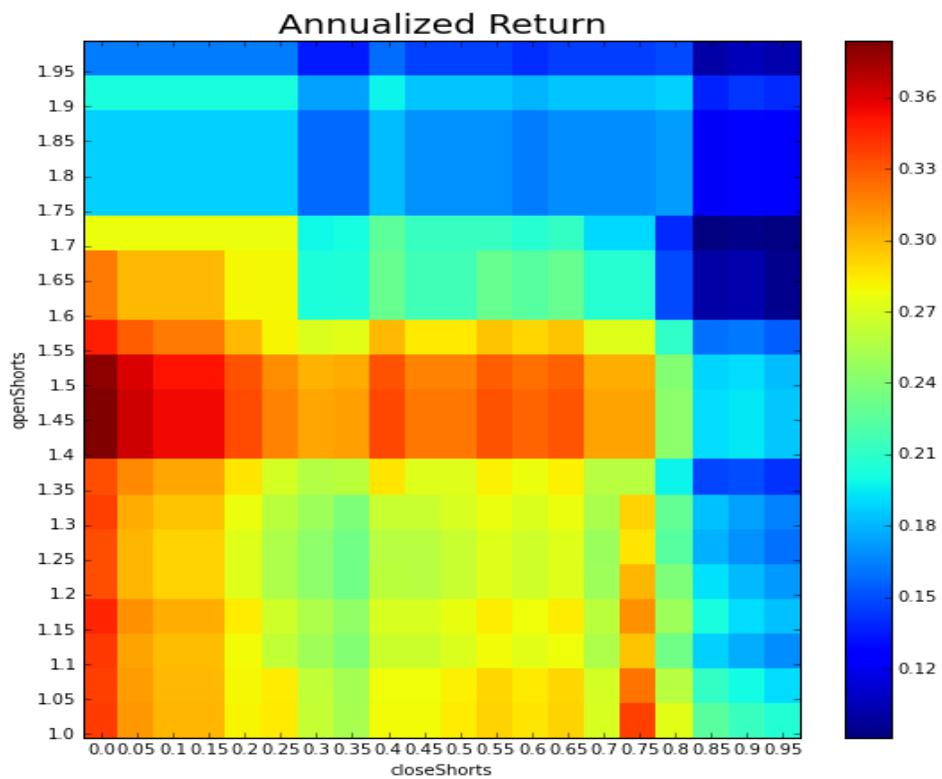

Figure 4

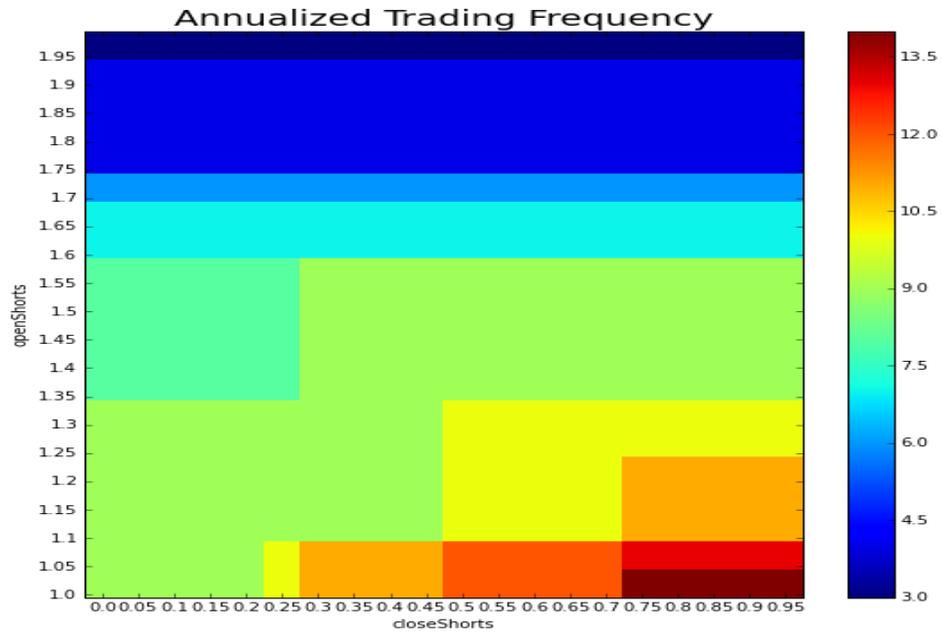

Figure 5

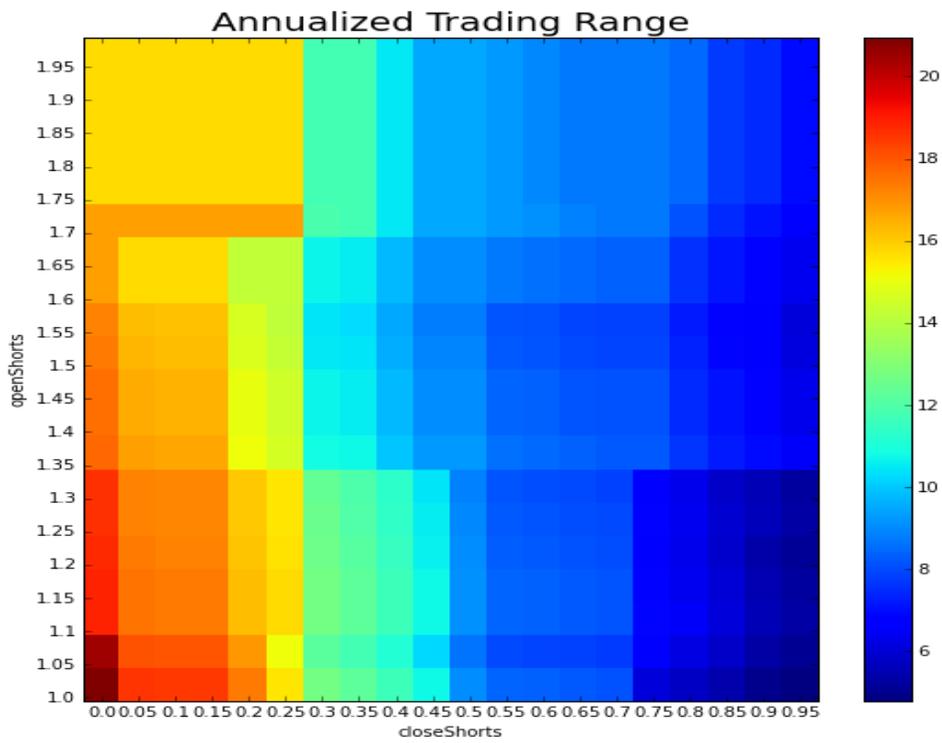

Figure 6

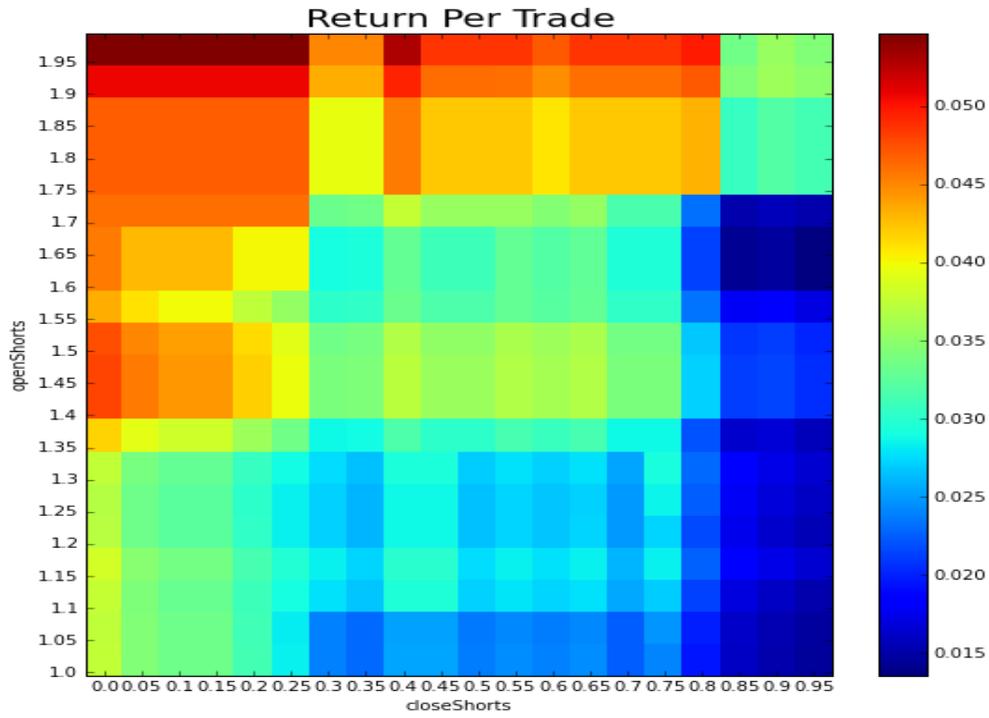

Figure 7

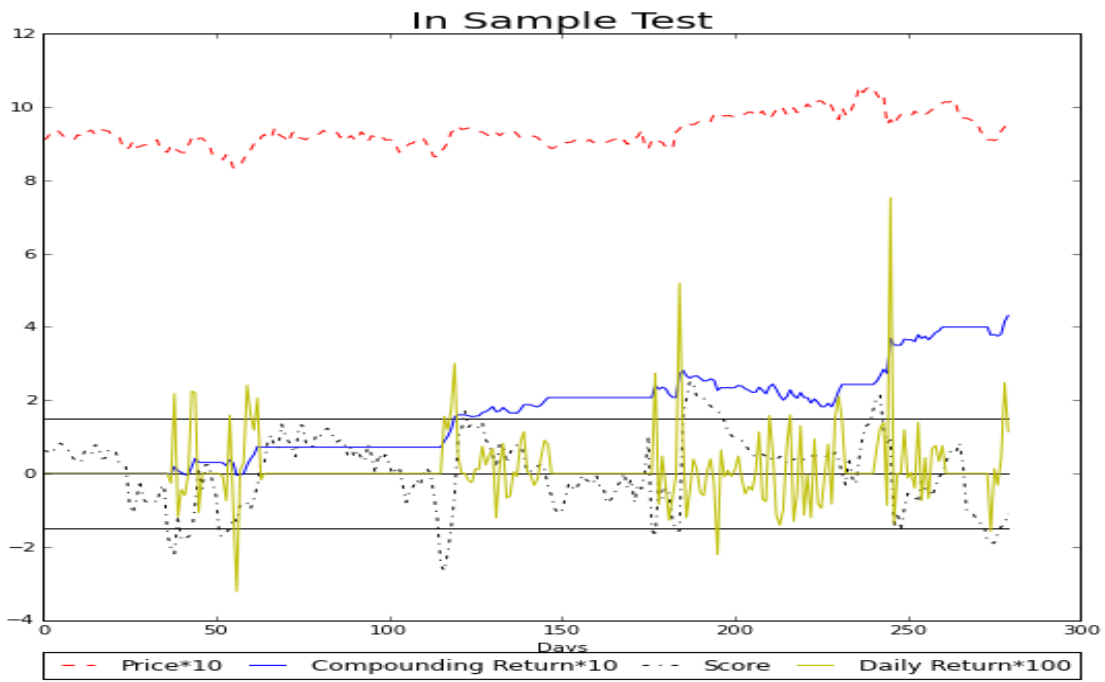

Figure 8

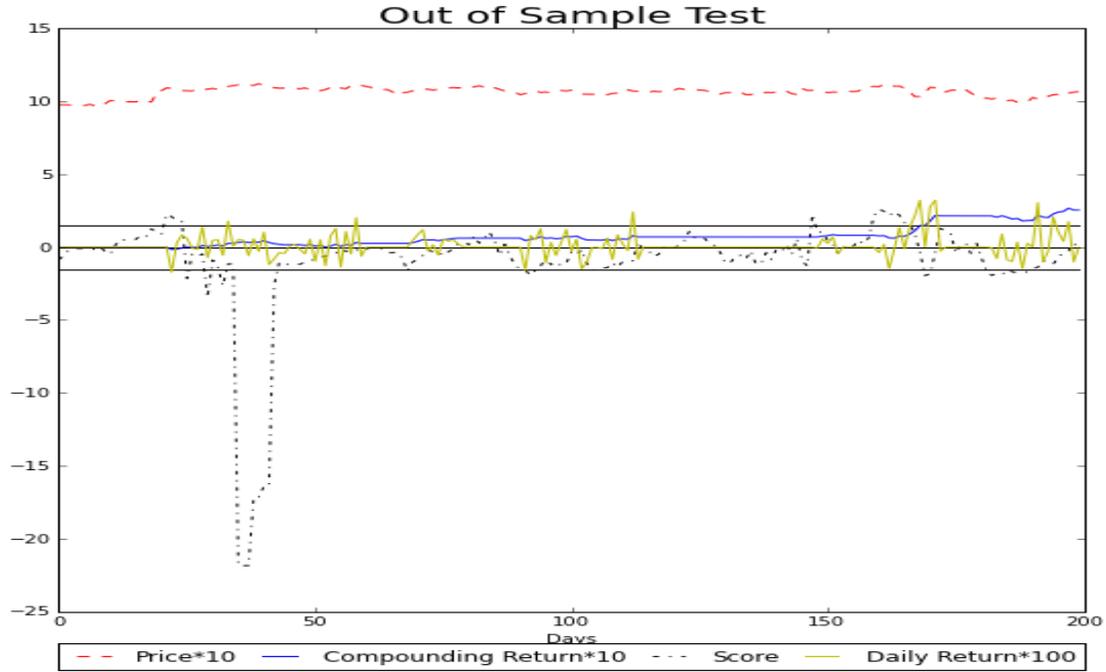

Figure 9

## 3.2 CCI vs O

We constructed a portfolio by holding $1 Crown Castle International Corp. CCI and shorting $0.367 Realty Income Corporation (O). The pair of stocks are selected from REITs sector. Similarly, results are shown in the following figures. The best threshold for this pair is $S_o = 1.15, S_c = 0.05$.

|  | Price | $\hat{\theta}$ | $\hat{\mu}$ | $\hat{\sigma}$ | $\hat{l}$ |
| --- | --- | --- | --- | --- | --- |
| CCI-O | empirical | 0.7204 | 10.7316 | 0.1960 | 2.9963 |
|  | simulated | 0.7114 | 12.9782 | 0.1909 | 3.0276 |

Table 3

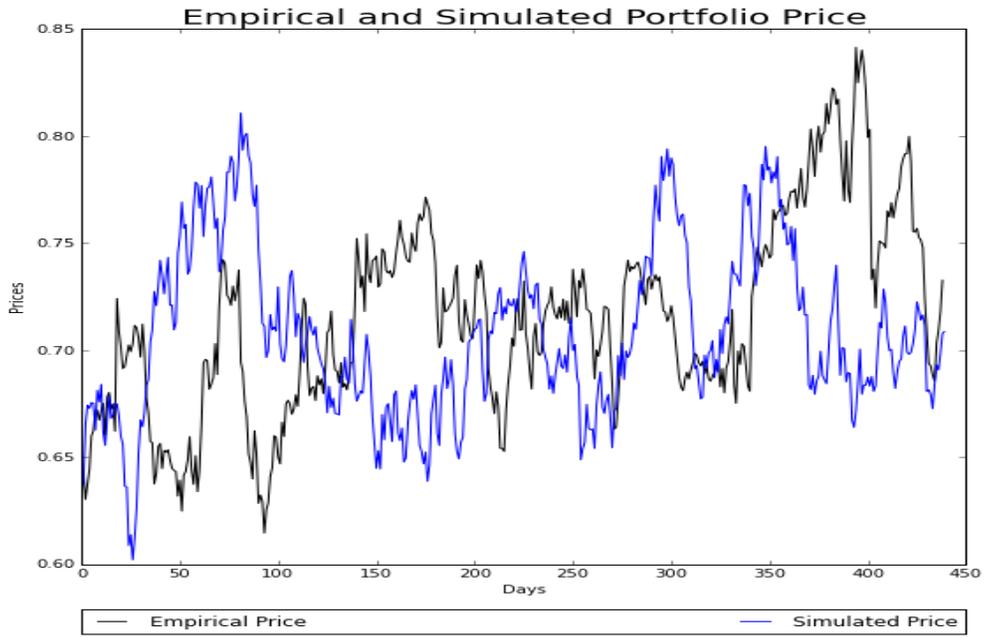

Figure 10

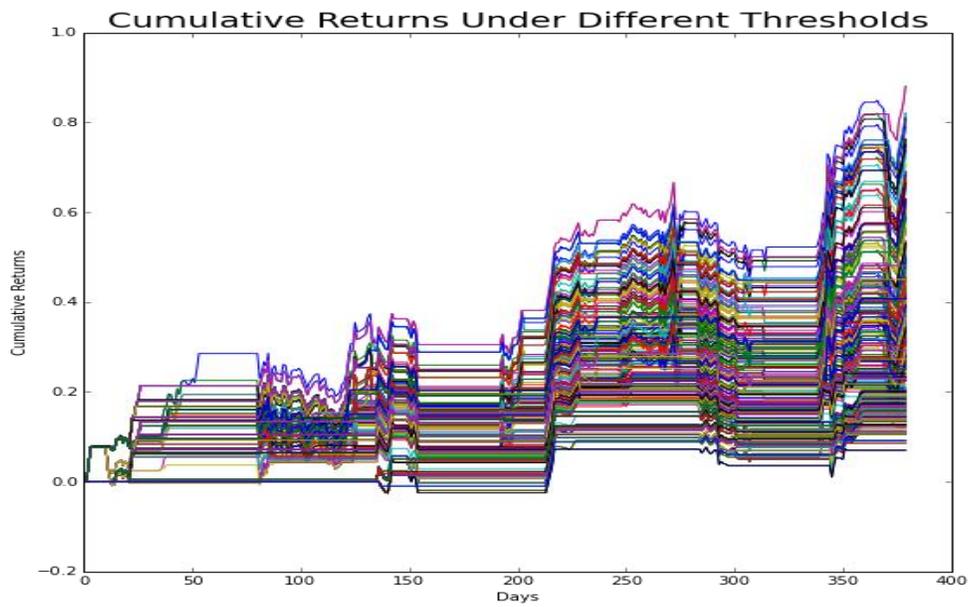

Figure 11

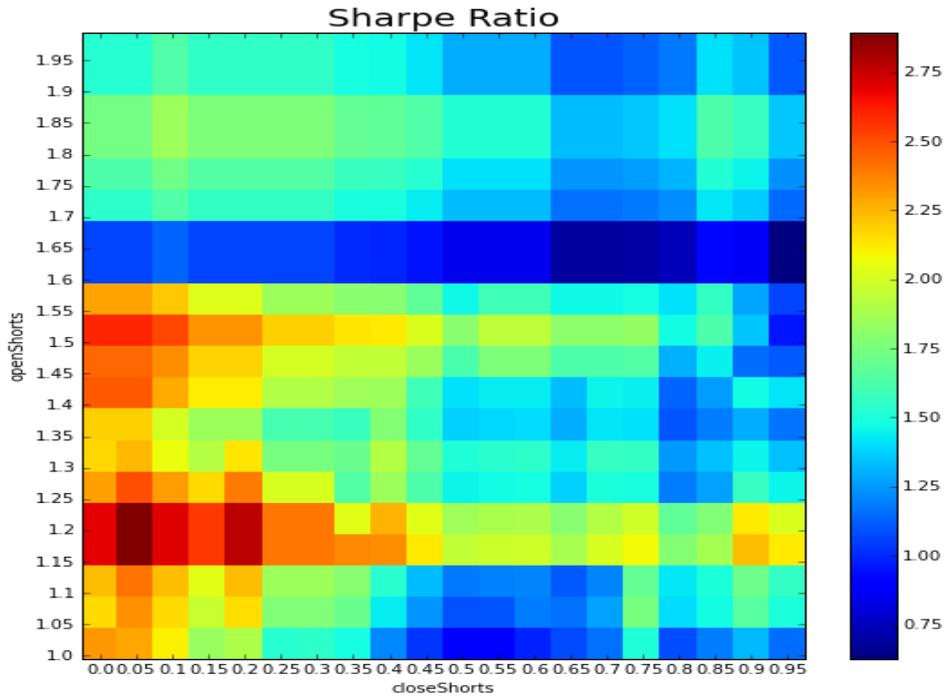

Figure 12

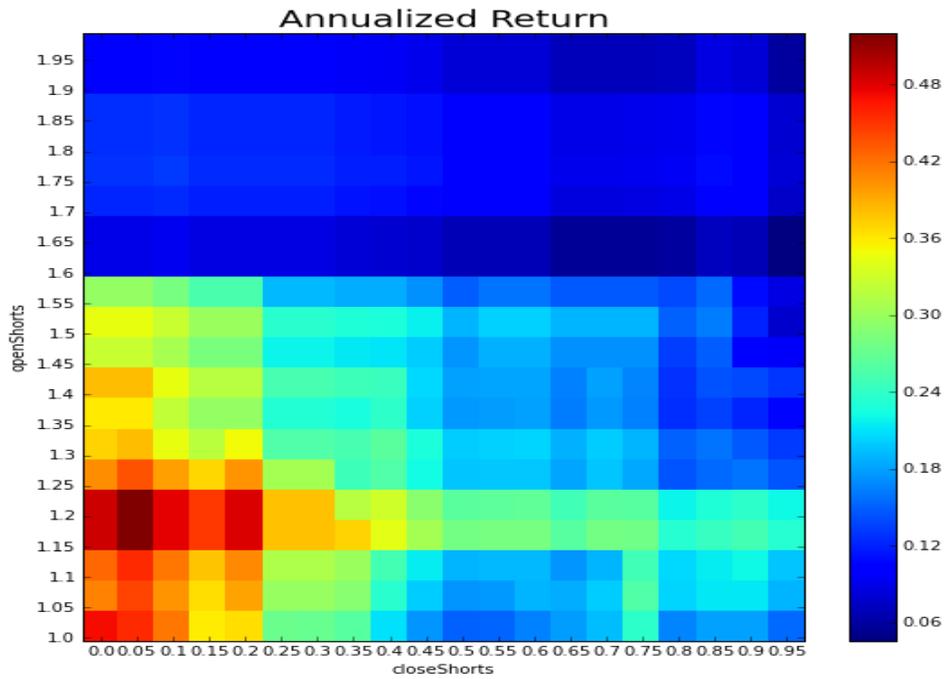

Figure 13

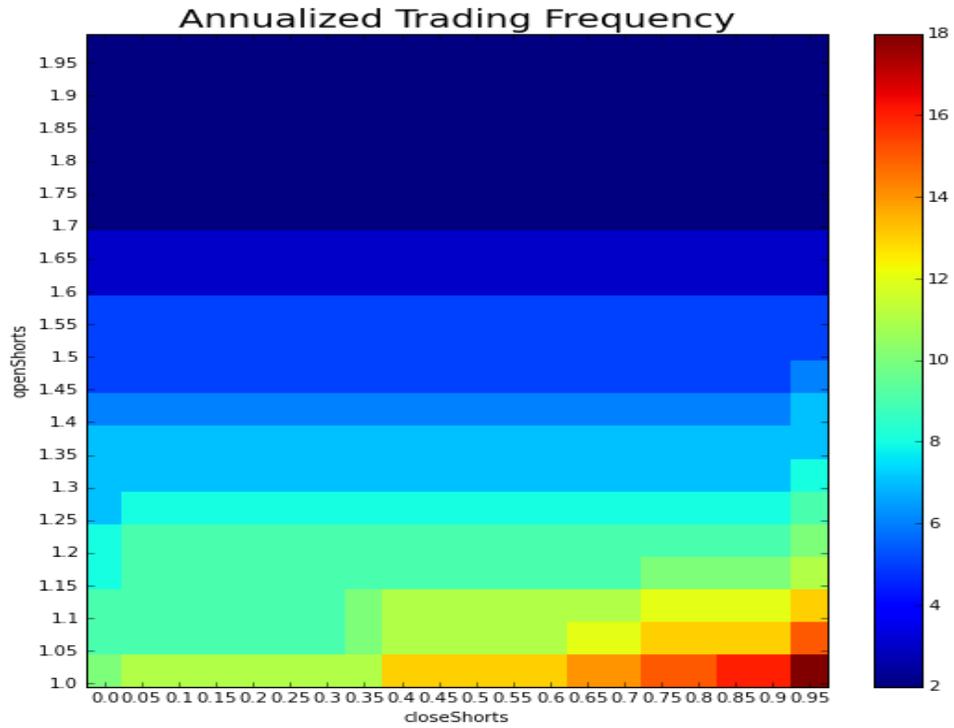

Figure 14

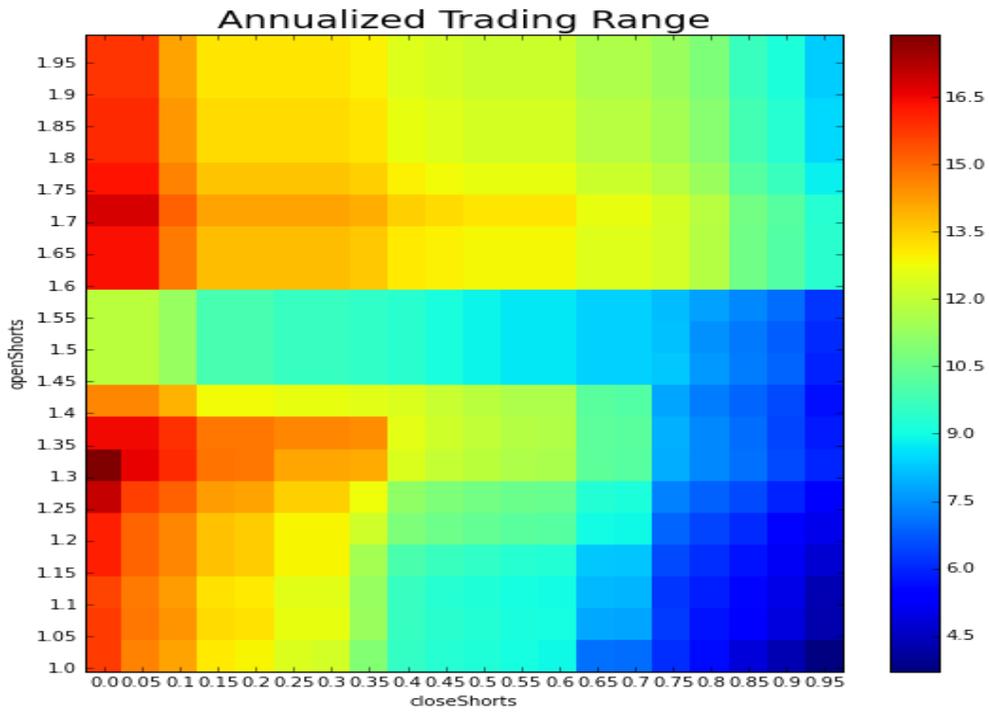

Figure 15

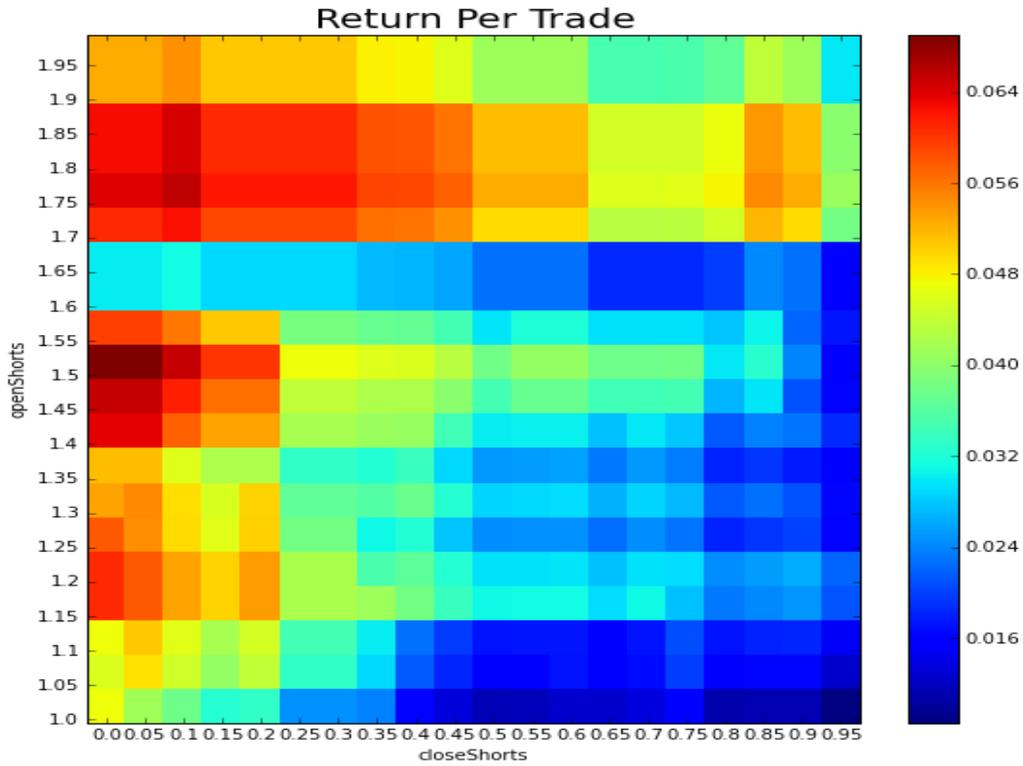

Figure 16

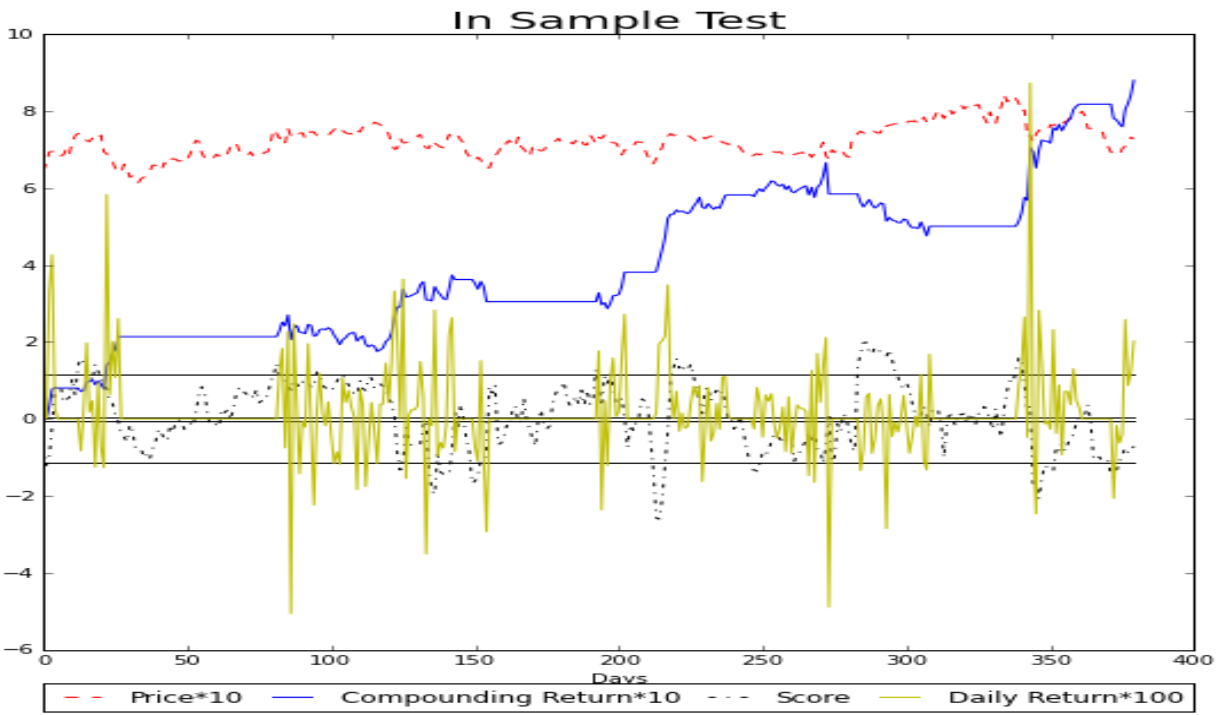

Figure 17

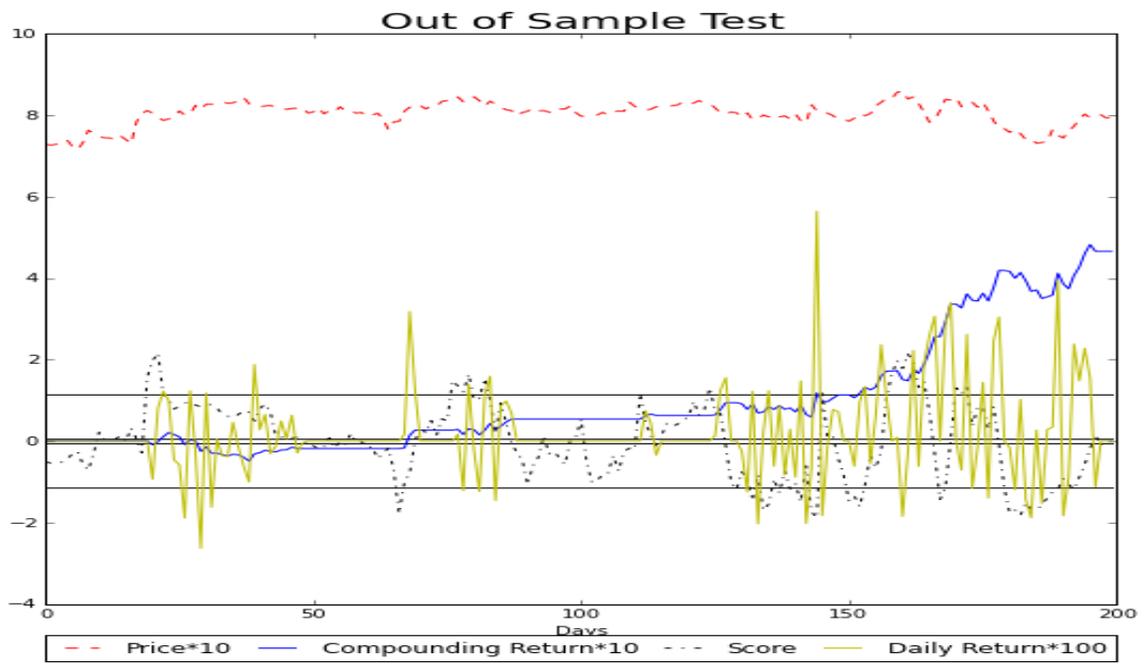

Figure 18